\documentclass{aa}
\usepackage{graphicx}
\begin{document}
\title{The Leo Triplet: common origin or late encounter?
\thanks{Based on observations collected with the 6m
telescope of the Special Astrophysical Observatory (SAO) of the
Russian Academy of Sciences (RAS) which is operated under the
financial support of Science Department of Russia (registration number
01-43) and on data from the Isaac Newton Group (ING) and Hubble Space
Telescope (HST) Archives.}}
\author {V. L. Afanasiev
\inst{1}
\and O. K. Sil'chenko
\inst{2,3,4}
}
\offprints{O. K. Sil'chenko, \email{olga@sai.msu.su}}
\institute{
Special Astrophysical Observatory, Nizhnij Arkhyz, 369167 Russia
\and
Sternberg Astronomical Institute, University av. 13,
Moscow 119992, Russia
\and
Isaac Newton Institute of Chile, Moscow Branch
\and
UK Astronomy Data Centre, Guest Investigator
}

\date{Received / Accepted }

\abstract{
The kinematics, structure, and stellar population properties
in the centers of two early-type spiral galaxies of the Leo Triplet,
NGC~3627 and NGC~3623, are studied by means of integral-field spectroscopy.
Unlike our previous targets, NGC~3384/NGC~3368 in the Leo~I Group and
NGC~5574/NGC~5576 in LGG~379, NGC~3623 and NGC~3627 do not appear
to experience a synchronous evolution. The mean ages of their circumnuclear
stellar populations are quite different, and the magnesium overabundance
of the nucleus in NGC~3627 is evidence for a very brief last star formation
event 1~Gyr ago whereas the evolution of the central part of NGC~3623
looks more quiescent. In the center of NGC~3627 we observe noticeable
gas radial motions, and the stars and the ionized gas in the center of 
NGC~3623 demonstrate more or less stable rotation. However, NGC~3623 has
a chemically distinct core -- a relic of a past star formation burst --
which is shaped as a compact, dynamically cold stellar disk with a radius
of $\sim 250-350$ pc which has been formed not later than 5 Gyr ago.
}

\authorrunning{Afanasiev \&\ Silchenko}
\titlerunning{Leo~3}

\maketitle

\section{Introduction}

It is a long-standing problem what mechanisms define the evolution of
galaxies, intrinsic or external. During the last years external mechanisms
have been winning the competition: modern cosmology prescribes multiple
mergers to be the main driving force of galactic evolution as regards
star formation process or morphological transformations. However,
intrinsic disk instabilities which may provoke a substantial secular
evolution are not left out entirely. From an observational point of view
one of the most promising approaches is to study in detail contemporary
properties of galaxies in different environments and to search for
signatures of environmental influence on these properties. Galaxy groups
constitute perhaps the least-studied type of environment, opposite to
clusters or pure field; meanwhile, their closeness to each other and moderate
relative velocities make them very suitable objects to demonstrate
the power of external mechanisms, of interactions in particular, to govern
the evolution of galaxies that are members of groups. So we aim 
to collect statistics on stellar population and structure properties in the
central parts of early-type galaxies in nearby groups.

We choose to compare the properties of galaxies of comparable luminositites
and of close morphological types, and this choice limits our sample of
galaxy groups: we need at least two early-type galaxies, close to each other
and bright enough, in each group. Up to now we have investigated stellar
population properties, the mean (luminosity-weighted) stellar age first
of all, in \object{NGC~5576}/\object{NGC~5574} (Sil'chenko \cite{sil97},
Sil'chenko et al. \cite{s0_3}) and
in \object{NGC~3379}/\object{NGC~3384}/\object{NGC~3368}
(Sil'chenko et al. \cite{we2003}). In both groups
we have detected a certain coherence of star formation events in the nuclei
of neighbouring galaxies. In the centers of the E-galaxy \object{NGC~5576} and
the S0-galaxy \object{NGC~5574} the iron-to-magnesium abundance ratio exceeds 
the solar value and the mean age of the stellar populations is 3~Gyr.
In the centers of the S0-galaxy \object{NGC~3384} and the Sab-galaxy
\object{NGC~3368} the mean age is also found to be 3 Gyr. Moreover,
in the centers of \object{NGC~3384} and
\object{NGC~3368} we have found decoupled kinematical substructures whose
spatial orientations match the orientation of the supergiant \ion{H}{I}
cloud (ring) encircling the central galaxies of the Leo~I group. We have
concluded that at least when a significant amount of intergalactic cool
gas is available, the circumnuclear evolution of the group member galaxies
have been strongly affected by their mutual interaction and/or by their
interaction with intergalactic gas clouds.

\begin{table}
\caption[ ] {Global parameters of the galaxies}
\begin{flushleft}
\begin{tabular}{lcc}
\hline\noalign{\smallskip}
NGC & 3623 & 3627  \\
\hline
Type (NED$^1$) & SAB(rs)a & SAB(s)b \\
$B_T^0$ (RC3$^2$) & 9.65 & 9.13  \\
Distance$^3$, Mpc  & \multicolumn{2}{c}{9.1} \\
$M_B$ (from $B_T^0$, RC3) & --20.15 & --20.67 \\
$R_{25}$, kpc (from $D_{25}$, RC3) &
     12.9 & 12.1 \\
$(B-V)_T^0$ (RC3) & 0.77 & 0.60  \\
$(U-B)_T^0$ (RC3) & 0.32 & 0.11 \\
$V_r$, $\mbox{km} \cdot \mbox{s}^{-1}$ (NED) & 807 &
      727  \\
{\it i}$_{\mbox{phot}}$$^4$  & $70.9^{\circ}$ & $64.2^{\circ}$  \\
{\it PA}$_{\mbox{phot}}$ (RC3) & $174^{\circ}$ & $173^{\circ}$  \\
$V_{\mbox{rot}}$, $\mbox{km} \cdot \mbox{s}^{-1}$, from HI$^4$ & 266 & 202 \\
$\sigma _*$, $\mbox{km} \cdot \mbox{s}^{-1}$ & 149$^5$ & 117$^6$ \\
$M_{\mbox{HI}}$$^7$, $10^8\,M_{\odot}$ & 2.7 & 7.6 \\
$M_{\mbox{H}_2}$$^8$, $10^8\,M_{\odot}$ & 3.7 & 32.3 \\
\hline
\multicolumn{3}{l}{$^1$\rule{0pt}{11pt}\footnotesize
NASA/IPAC Extragalactic Database}\\
\multicolumn{3}{l}{$^2$\rule{0pt}{11pt}\footnotesize
Third Reference Catalogue of Bright Galaxies}\\
\multicolumn{3}{l}{$^3$\rule{0pt}{11pt}\footnotesize
Paturel et al. (\cite{patdist})}\\
\multicolumn{3}{l}{$^4$\rule{0pt}{11pt}\footnotesize
Haynes(\cite{haynes81})}\\
\multicolumn{3}{l}{$^5$\rule{0pt}{11pt}\footnotesize
Whitmore et al.(\cite{whit79})}\\
\multicolumn{3}{l}{$^6$\rule{0pt}{11pt}\footnotesize
Heraudeau \&\ Simien(\cite{hs98})}\\
\multicolumn{3}{l}{$^7$\rule{0pt}{11pt}\footnotesize
Haynes et al.(\cite{haynes79})}\\
\multicolumn{3}{l}{$^8$\rule{0pt}{11pt}\footnotesize
Sage (\cite{sage}), corrected for the new distance}
\end{tabular}
\end{flushleft}
\end{table}

In this work we consider two nearby early-type spiral galaxies belonging
to the famous \object{Leo Triplet}, \object{NGC~3623} and \object{NGC~3627}.
The third member of this company, \object{NGC~3628}, is of rather late type,
and moreover, it is seen edge-on and has a dust lane completely obscuring
the nucleus. The main global properties of \object{NGC~3623} and
\object{NGC~3627} are given in Table~1. The projected
distance between the two galaxies is only $20\arcmin$, or
$\sim 50$ kpc; it is comparable to the distance between our Galaxy and the
LMC, and considering the existence of the Magellanic Stream, we would
expect tidal interaction between \object{NGC~3623} and \object{NGC~3627}.
However these two galaxies are always mentioned as `non-interacting pair'
(e.g. RC1 or RC2); instead violent interaction has been supposed to exist 
between \object{NGC~3627} and \object{NGC~3628}, starting from the works of
Zwicky (\cite{zwicky}) and Vorontsov-Velyaminov (\cite{vv_p1}),
though the distance between
\object{NGC~3627} and \object{NGC~3628} is almost twice that between
\object{NGC~3627} and \object{NGC~3623} and the difference in systemic
velocities is the same (\object{NGC~3623} and \object{NGC~3628}
have similar $v_{\mbox{LSR}}$). By inspecting either Table~1, or a DSS
image of the Leo Triplet, it can be noted that \object{NGC~3623} and
\object{NGC~3627} look almost like twins: they have the same sizes, close
morphological types, close spatial orientations (their lines of nodes
are parallel), the same sense of rotation with the southern parts of both
receding, close luminosities, etc. \object{NGC~3623} is a little more
massive, if witness its higher rotation velocity and stellar velocity
dispersion, and \object{NGC~3627} is more gas-rich with consequently more
intense global star formation and bluer integrated colours. Why don't they 
influence each other? Both galaxies have very similar global bars,
with radii of $R\sim 2$ kpc, that are aligned and so may be of tidal 
origin. Could we perhaps find signatures of coherent evolution in
the central parts of the galaxies? Here we present results of
2D spectroscopy of the circumnuclear regions of \object{NGC~3623} and
\object{NGC~3627}.
The layout of the paper is as follows. We report our observations and
other data which we use in Sect.~2. The radial variations of the
stellar population properties are analysed in Sect.~3, and
in Sect.~4 two-dimensional velocity fields obtained
by means of 2D spectroscopy for the central parts of \object{NGC~3623}
and \object{NGC~3627} are presented. Sect.~5 gives discussion and our
conclusions.

\section{Observations and data reduction}

The spectral data which we analyse in this work are obtained with
two different integral-field spectrographs: the fiber-lens
Multi-Pupil Fiber Spectrograph (MPFS) at the 6m telescope of the Special
Astrophysical Observatory of the Russian Academy of Sciences (SAO RAS)
and the international Tigre-mode SAURON at the 4.2m William
Herschel Telescope at La Palma.

The last modification of the MPFS became operational
in the prime focus of the 6m telescope in 1998
(http://www.sao.ru/hq/lsfvo/devices/mpfs/).
It is a fiber-lens system: densely packed square microlenses
in the focal plane of the telescope create a set of micropupils,
and the fibers after them transmit the light from $16\times 15$ square
elements of the galaxy image to the slit of the spectrograph together
with 16 additional fibers that transmit the sky background light taken 
at a distance of $4^{\prime}$ from the galaxy, so separate sky exposures 
are not necessary. The size of one spatial element was approximately
$1\arcsec \times 1\arcsec$; a CCD TK $1024 \times 1024$ detector
was used. The reciprocal dispersion was 1.35~\AA\ per pixel, with
a spectral resolution of 5~\AA, rather stable over the field of view.
The wavelength calibration was done by exposing a spectrum
of a hollow cathode lamp filled with helium, neon, and argon
before and after the galaxy exposures;
the internal accuracy of linearization was typically 0.25~\AA\ in the
green and 0.1~\AA\ in the red; additionally we checked the accuracy
and the absence of a systematic velocity shift by measuring strong emission
lines of the night sky [\ion{O}{I}]$\lambda$5577 and
[\ion{O}{I}]$\lambda$6300.
We obtained MPFS data in two spectral ranges,
green, 4300--5600~\AA, and red, 5900--7200~\AA.
The green spectra were used to calculate
the Lick indices H$\beta$, Mgb, Fe5270, and Fe5335 which are suitable
for determining metallicity, age, and Mg/Fe ratio of old stellar populations
(Worthey \cite{worth94}), and they are also used for cross-correlation
with the spectrum of a template star, usually of G8III--K1III spectral type,
in order to obtain the line-of-sight velocity field for the stellar
component and a map of the stellar velocity dispersion. The red spectral range
contains strong emission lines H$\alpha$ and [\ion{N}{II}]$\lambda$6583
and is used to derive line-of-sight velocity fields for the ionized gas.
To calibrate the new MPFS index system onto the standard Lick one,
we have observed 15 stars from the list of Worthey et al.(\cite{woretal})
during four observational runs and calculated the
linear regression formulae to transform our index measurements onto the
Lick system; the rms scatters of points near the linear
dependencies are about 0.2~\AA\ for all 4 indices under consideration,
so they are within the observational errors of Worthey et al. (\cite{woretal}).
To correct the index measurements for the stellar velocity dispersion
which is usually substantially non-zero in the centers of early-type
galaxies, we have applied smoothing to the spectrum of the standard
star \object{HD~97907} by a set of Gaussians of various widths; the derived
dependencies of the index corrections on $\sigma$ were approximated by
polynomials of the 4th order and applied to the measured index values
before they are put onto the Lick system.

The second 2D spectrograph from which we used the data is a new
instrument, SAURON, operated at the 4.2m William Herschel Telescope (WHT)
on La Palma -- for its detailed description see Bacon et al.(\cite{betal01})
 and for some preliminary scientific results see de Zeeuw et al.
(\cite{sau2}).
We have taken the data for \object{NGC~3623} observed in March 2000
from the open ING Archive of the UK Astronomy Data Centre. Briefly,
the field of view of this instrument is $41\arcsec \times 33\arcsec$
with a spatial element size of $0\farcs 94 \times 0\farcs 94$.
The sky background is taken 2 arcminutes from the center of the galaxy
and is exposed simultaneously with the target. The fixed spectral range is
4800-5400~\AA, the reciprocal dispersion is 1.11~\AA\--1.21~\AA,
varying from the left to the right edge of the frame. The comparison
spectrum is that of neon, and we made the linearization by fitting a polynomial
of the 2nd order with an accuracy of 0.07~\AA. The index system is
checked by using stars from the list of Worthey et al. (\cite{woretal})
that have been observed during the same observational run. The regressions
describing the index system of the March 2000 observational run
when \object{NGC~3623} was observed can be found in our paper
on the Leo~I group where the SAURON
data for \object{NGC~3384} are analysed (Sil'chenko et al. \cite{we2003}).
The relations between instrumental and standard-system indices
were approximated by linear formulae which were applied to
our measurements to put them onto the standard Lick system.
The stellar velocity dispersion effect was corrected in the
same manner as for the MPFS data.

The full list of the exposures made for \object{NGC~3623} and \object{NGC~3627}
with two 2D spectrographs is given in Table~2.

\begin{table*}
\caption[ ] {2D spectroscopy of the galaxies studied}
\begin{flushleft}
\begin{tabular}{lllllcc}
\hline\noalign{\smallskip}
Date & Galaxy & Exposure & Configuration & Field
& Spectral range & Seeing \\
\hline\noalign{\smallskip}
9 Feb 2000 & NGC~3623 & 45 min & BTA/MPFS+CCD $1024 \times 1024$ &
$16\arcsec \times 15\arcsec $ & 4200-5600~\AA\ & $1\farcs 4$ \\
12 Feb 2000 & NGC~3627 & 45 min & BTA/MPFS+CCD $1024 \times 1024$ &
$16\arcsec \times 15\arcsec $ & 4200-5600~\AA\ & $3\farcs 3$ \\
14 Feb 2000 & NGC~3623 & 52 min & BTA/MPFS+CCD $1024 \times 1024 $ &
$16\arcsec \times 15\arcsec $ & 5800-7200~\AA\ & $3\farcs 5$ \\
15 Feb 2000 & NGC~3627 &  30 min & BTA/MPFS+CCD $1024 \times 1024 $ &
$16\arcsec \times 15\arcsec $ & 5800-7200~\AA\ & $3\farcs 5$ \\
29 Mar 2000 & NGC~3623, Pos.1 & 120 min & WHT/SAURON+CCD $2k\times 4k$ &
$33\arcsec\times 41\arcsec$ & 4800-5400~\AA\ & $1\farcs 5$ \\
30 Mar 2000 & NGC~3623, Pos.2 & 120 min & WHT/SAURON+CCD $2k\times 4k$ &
$33\arcsec\times 41\arcsec$ & 4800-5400~\AA\ & $1\farcs 6$ \\
\hline
\end{tabular}
\end{flushleft}
\end{table*}

Additionally, to know the central structure of the galaxies,
we have retrieved HST data from the HST Archive.
Since both galaxies are dust-rich in the center, we try to use
the reddest images available. \object{NGC~3623} was exposed through
the F814W filter in the frame of the program
of S. Smartt (ID 9042) of searching progenitors of SNe (the nucleus
of the galaxy is located in the WF3-frame, so the scale is $0\farcs 1$ 
per pixel), and \object{NGC~3627} was observed in the frame of the programs
of J. Mulchaey on the fueling of active nuclei (ID 7330) with NICMOS2
and of W. Sparks' snapshots (ID 7919) with NICMOS3.
To probe the large-scale structures of the galaxies under consideration, 
we have also retrieved the 2MASS images from NED for both galaxies.

All the data, spectral and photometric, except the data obtained with
the MPFS, have been reduced with the software produced by Dr. V.V. Vlasyuk
of the Special Astrophysical Observatory (Vlasyuk \cite{vlas}).
Primary reduction
of the data obtained with the MPFS was done in IDL with a software
created by one of us (V.L.A.). The Lick indices were calculated with
our own FORTRAN program as well as using the FORTRAN program of
Dr. A. Vazdekis which also provides the index error calculation.

\section{Stellar population properties in the centers of NGC 3623 and NGC 3627}

Since the determination of stellar population properties requires very high 
accuracy of the Lick indices, we 
use two-dimensional maps for a morphological analysis of stellar
population substructures, and azimuthally averaged profiles for
quantitative age and abundance estimations.

\subsection{NGC 3623}

\begin{figure*}
\centering
  \includegraphics[width=17cm]{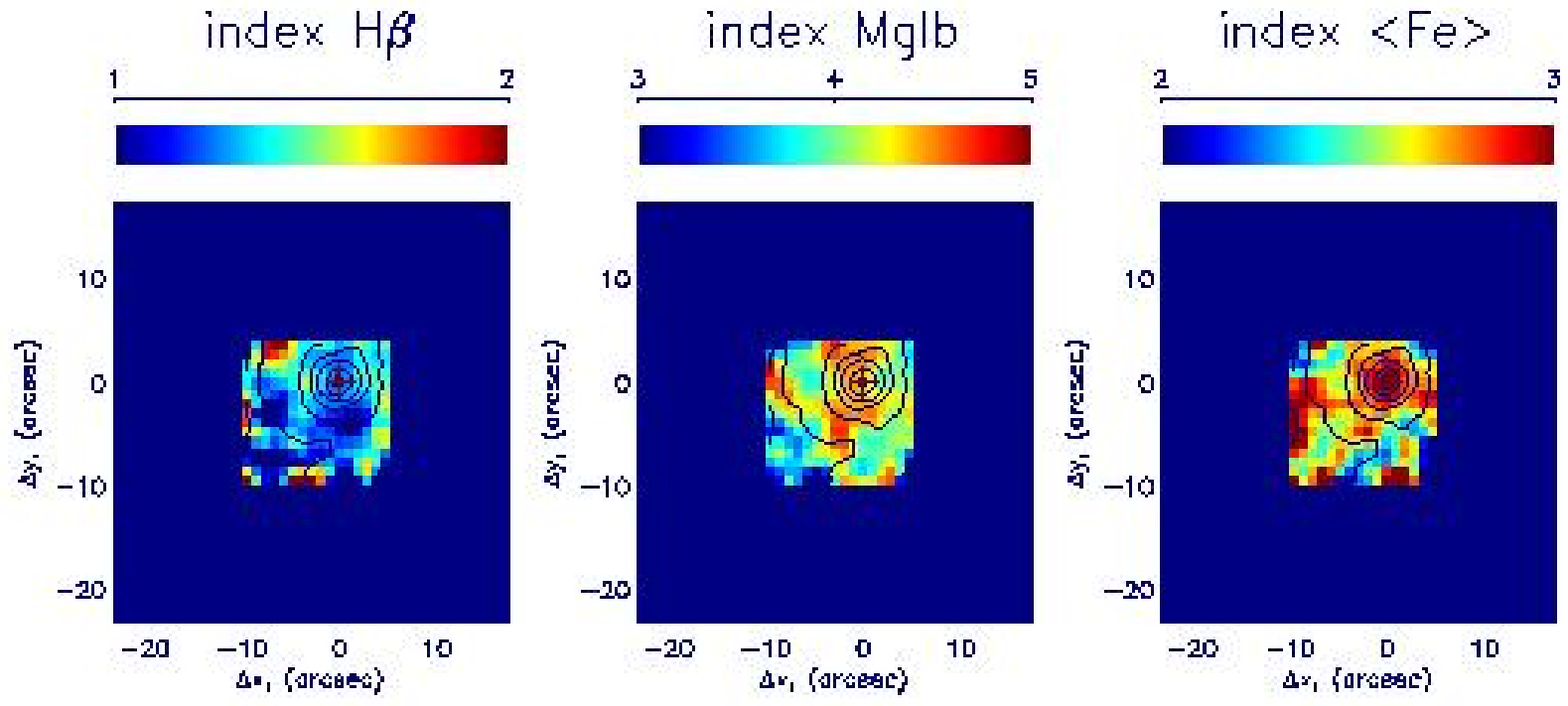}
  \caption{The MPFS index maps for NGC 3623;
  $<\mbox{Fe}> \equiv$(Fe5270+Fe5335)/2. The green
  ($\lambda$5000~\AA ) continuum is overlaid by isophotes
   marking 10\%, 20\%, 30\%, etc, of the maximum surface
   brightness.}
  \label{mpfs3623}
\end{figure*}

\begin{figure*}
\centering
  \includegraphics[width=17cm]{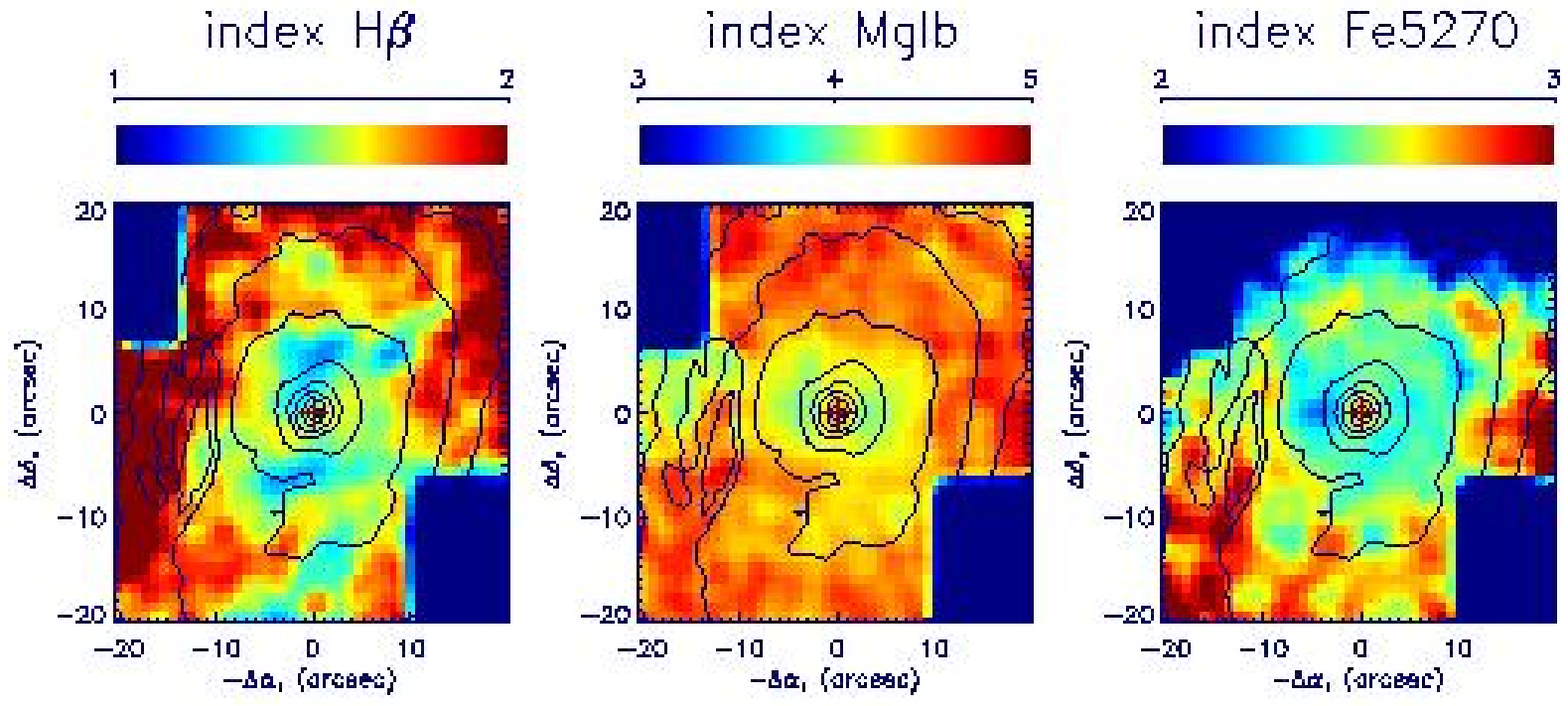}
  \caption{The SAURON index maps for NGC 3623. The green
  ($\lambda$5000~\AA ) continuum is overlaid by isophotes. }
  \label{sau3623}
\end{figure*}

Figure~1 presents three $16\arcsec \times 15\arcsec$ maps of the Lick indices,
H$\beta$, Mgb, and $\langle \mbox{Fe} \rangle \equiv$(Fe5270+Fe5335)/2,
constructed from the MPFS data, and Fig.~2 presents three
$40\arcsec \times 40\arcsec$ maps of the Lick indices H$\beta$, Mgb, and
Fe5270 constructed from the SAURON data. The latter maps are obtained by
combining two independent data sets, Pos.I and Pos.II (see Table~2), which
are exposures of
the northern and the south-eastern parts of the galaxy, though both
include the nucleus. Although the sizes of the fields of view differ greatly 
for the MPFS and SAURON, both Fig.~1 and Fig.~2  show the main 
characteristic feature of the stellar population distribution: a chemically
decoupled  core of \object{NGC~3623}  which looks like a resolved
structure enhanced in the magnesium index and elongated north-south. 
Previously Proctor et al.
(\cite{prsanr00}) and Proctor (\cite{prdiss}) reported a Mg-index peak
in the nucleus of \object{NGC~3623} from their long-slit observations
along the minor axis. Proctor (\cite{prdiss})  even gives 
a quantitative estimate of the metallicity drop between the nucleus, unresolved
in their observations, and the bulge: from $+0.6$ dex to $\sim +0.1$ dex, or
by a factor of 3. But now for the first time we can establish the morphology
of the chemically decoupled entity: it is 
a circumnuclear disk or a bar with a radius of $\sim 8\arcsec$ 
($\sim 350$ pc), almost aligned with the global
line of nodes of the galaxy. From Fig.~1 one may get an impression that the
distribution of $\langle \mbox{Fe} \rangle$ shows a much more compact
peak than that of Mgb; it may signify that the characteristic time of
secondary star formation that has produced the chemically decoupled 
substructure varied along the radius. The maps of Fig.~2, with their large 
extension, reveal two further interesting features of the index distributions:
the Mgb index reaches a minimum in a shallow, almost circular (spherical?) 
area that encircles the metal-enriched disk (or bar), and the radius of this 
area is $\sim 10\arcsec$. The most natural explanation -- that we see an old, 
rather metal-poor bulge -- is in some contradiction to the known estimates 
of the size of the bulge of NGC~3623
($18\arcsec$, Hogg et al. \cite{hogg}; $20\arcsec$, Burkhead \&\ Hutter
\cite{burkphot}).
The other interesting feature, two symmetric H$\beta$ minima at $5\arcsec -
8\arcsec$ from the center located near the major axis of the isophotes 
(and of the Mgb isolines), may be a consequence of H$\beta$ index contamination 
by gas emission. In this case, we would have detected an ionized-gas torus 
edge-on; and if so, the circumnuclear chemically decoupled substructure would 
have a high probability of being a bar.

\begin{figure}
 \resizebox{\hsize}{!}{\includegraphics{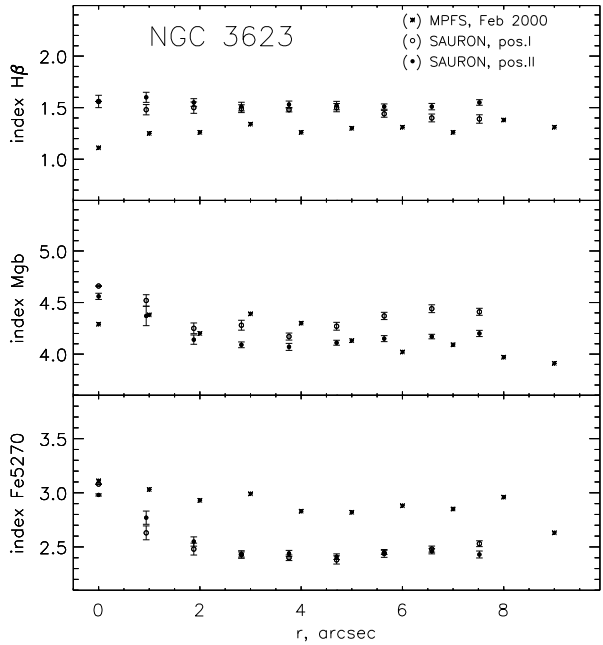}}
 \caption{The MPFS azimuthally-averaged index profiles for NGC 3623
 in comparison with the two sets of the SAURON data.} 
 \label{indcomp1}
\end{figure}

\begin{figure}
 \resizebox{\hsize}{!}{\includegraphics{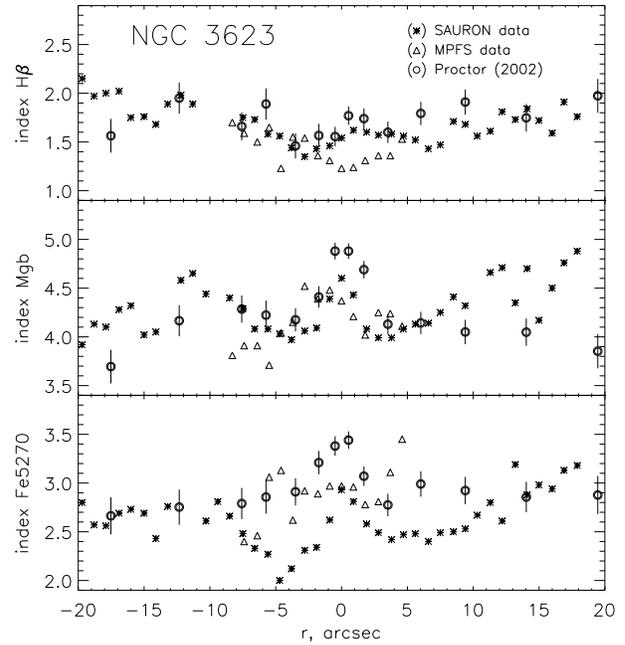}}
 \caption{The minor-axis long-slit index cross-sections by Proctor (2002) 
 for NGC 3623 are plotted in comparison with the minor-axis index profiles 
 simulated from the MPFS 2D index maps and from the SAURON 2D index maps
 with a digital slit width of 1.25\arcsec.} 
 \label{indcomp2}
\end{figure}

Since for \object{NGC~3623} we have two sources of data, we have to compare
them and to select basic data for quantitative estimates of the stellar
population properties. Figures~3 and 4 present this comparison: in Fig.~3
we compare azimuthally averaged index profiles according to the MPFS data
and to two independent SAURON data sets, and in Fig.~4 we compare
minor-axis index cross-sections simulated over the 2D index maps
with the long-slit data of Proctor (\cite{prdiss}).
As to Fig.~3, the two SAURON data sets
agree perfectly despite the very different positions of the galactic
nucleus in the micropupil frame and the resulting slightly different spectral
response and resolution. The magnesium data of the MPFS and of SAURON
show no noticeable discrepancy but the other two indices do: 
the MPFS H$\beta$ is lower than the SAURON one by some
0.2--0.3~\AA, and on the contrary, the SAURON Fe5270 is lower than the MPFS
estimates by 0.4--0.5~\AA. The same shifts were found by us for the index
measurements in \object{NGC~3384} (Sil'chenko et al. \cite{we2003});
and since \object{NGC~3384} was observed with SAURON during the same 
observational run of March/April 2000, we conclude that these shifts 
 may be systematic. The comparison of the
SAURON data with the independently calibrated data of Proctor (\cite{prdiss})
in Fig.~4 assures us that the SAURON H$\beta$ estimates are correct
whereas the Fe5270 estimates in the central part of the galaxy are too low,
by the same 0.4--0.5~\AA. The same conclusions have been made by
Sil'chenko et al. (\cite{we2003}) for \object{NGC~3384}.  Moreover,
we have also found a shift of the MPFS H$\beta$ index profiles in the same 
sense for a dozen other galaxies in common with the data of
Proctor (\cite{prdiss}) and/or Fisher et al. (\cite{fisher96}) whereas our
iron indices are quite consistent with their measurements. Considering all
these effects, for safety we take for \object{NGC 3623} the 
H$\beta$ estimates from the SAURON data and the $\langle \mbox{Fe} \rangle $
estimates from the MPFS data to make a quantitative analysis of the 
 luminosity-weighted properties of the stellar population by comparing 
the observed Lick indices to evolutionary synthesis models.

\subsection{NGC 3627}

\begin{figure*}
\centering
  \includegraphics[width=17cm]{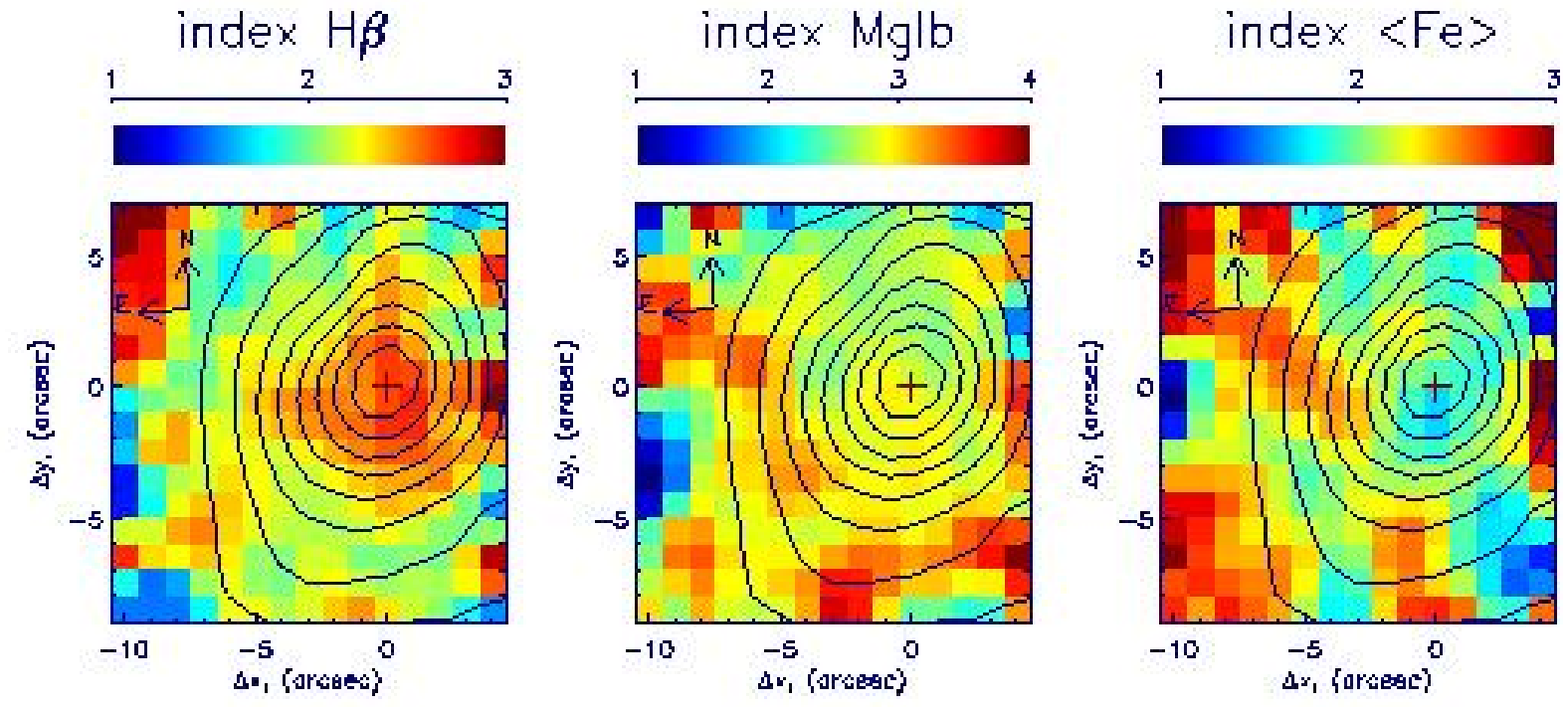}
  \caption{The MPFS index maps for NGC 3627;
  $<\mbox{Fe}> \equiv$(Fe5270+Fe5335)/2. The green
  ($\lambda$5000~\AA ) continuum is overlaid by isophotes.}
  \label{mpfs3627}
\end{figure*}

Figure~5 presents the index maps obtained for this galaxy with the MPFS.
They differ substantially from the index maps of \object{NGC~3623} (Fig.~1).
The magnesium map looks rather flat, but the maps of H$\beta$ and
$\langle \mbox{Fe} \rangle$ do not: H$\beta$ has an unresolved maximum and
$\langle \mbox{Fe} \rangle$ has an unresolved minimum near the center.
Both extrema are shifted by $1\arcsec - 1\farcs 5$ to the south-south-west
from the continuum peak; as the shifts are identical and the isophotes
are slightly distorted, we suggest that perhaps the H$\beta$ and
continuum peaks do not coincide due to a dust effect. So we should treat
the substructure indicated by the high H$\beta$ and low
$\langle \mbox{Fe} \rangle$ as the nucleus. Therefore, inspecting
the maps of all three indices together, we can immediately suggest that
the nucleus of \object{NGC~3627} is considerably younger than the surrounding 
bulge and that it is distinguished by an increased magnesium-to-iron ratio.

\subsection{Abundances and ages}

\begin{figure}
 \resizebox{\hsize}{!}{\includegraphics{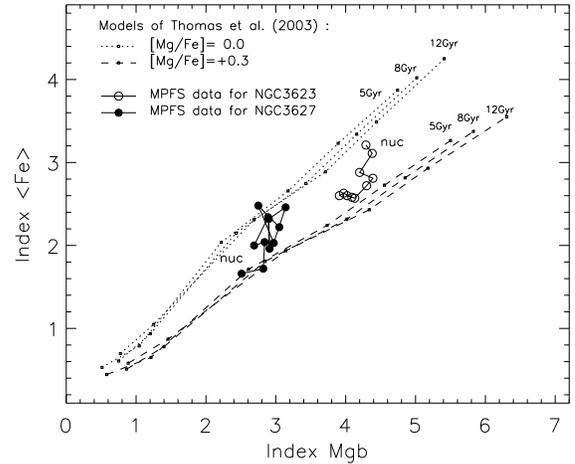}}
 \caption{`Index-index' $<\mbox{Fe}>$ vs Mgb diagnostic diagram
 for the azimuthally averaged  Lick indices in the center of NGC~3623 and
 NGC~3627 taken along the radius with a step of 1\arcsec\ (signs connected
 by solid lines, the nuclei are marked by `nuc').
 The models of Thomas et al. (2003) for [Mg/Fe]=0.0 and +0.3
 are plotted for reference frame; the small signs connected by
 pointed and dashed lines represent stellar population 
 models of equal ages; the metallicities for the models are
 +0.67, +0.35, 0.0, -0.33, -1.35, and -2.25 if one takes the signs from
 top to bottom.}
 \label{mgfe}
\end{figure}

Figure~6 presents a $\langle \mbox{Fe} \rangle$ {\bf vs} Mgb diagram.
Besides the azimuthally averaged MPFS data for \object{NGC~3623} and
\object{NGC~3627}, we have also plotted the most modern evolutionary synthesis
 models for old simple stellar populations by Thomas et al. (\cite{thomod})
 calculated for two values of [Mg/Fe], 0.0 and $+0.3$.
The observational data points lie mostly between two model sets, so we
may conclude that both bulges, that of \object{NGC~3623} and that
of \object{NGC~3627}, have [Mg/Fe]$=+0.1-+0.2$. The nuclei seem to be
slightly off. As for the nucleus of \object{NGC~3623}, its
displacement toward the locus of [Mg/Fe]$=0.0$ may be explained by a lower
age; this explanation is favoured by the results of
Proctor (\cite{prdiss}) for \object{NGC~3623} who, by involving also
the Balmer absorption lines of higher orders, has found a minimum of
the stellar population age in the nucleus and an age maximum
at $R\sim 3\arcsec$ as well as a flat profile of [Mg/Fe] of about
 $+0.15$. But the nucleus of \object{NGC~3627} certainly has
[Mg/Fe]$\ge +0.3$ implying a brief last event of star formation. As for
the total metallicity, \object{NGC~3627} is more metal-poor than
\object{NGC~3623}. In particular, the nucleus of \object{NGC~3623}
is supersolar, [m/H]$\ge +0.3$ (Proctor, \cite{prdiss},
gives [m/H]$\approx +0.5$), and
the nucleus of \object{NGC~3627} is
subsolar, [m/H]$\sim -0.3$. Considering
the well-known relation between total luminosity of galaxies and their
metallicity, we must note that the close luminosities and morphologies of
the two galaxies (\object{NGC~3627} is even slightly more luminous,
see Table~1) have not made us expect the metallicity difference
by a factor of 4 between \object{NGC~3623} and \object{NGC~3627}
-- such a metallicity difference usually corresponds to a difference
of integrated absolute magnitudes of 4 mag.

To determine the ages of the circumnuclear stellar populations in
\object{NGC~3623} and \object{NGC~3627}, we have to compare the stellar
hydrogen-line H$\beta$ index to some metal-line index.
Here a severe problem exists in that the Balmer absorption H$\beta$ is
somewhat contaminated by emission. To correct the measured Lick 
H$\beta$ index for the emission H$\beta$ we use the well-known fact that
an H$\alpha$ emission line is always much stronger than H$\beta$ and
an H$\alpha$ absorption line is always weaker than H$\beta$, especially
in spectra of intermediate-age populations, so the equivalent width of
the emission line H$\alpha$ can be measured more precisely than that of
H$\beta$. In this work we use the results of Ho et al.(\cite{hofil3})
for the nuclei of \object{NGC~3623} and \object{NGC~3627}, because our
efforts to make a Gauss analysis of the red spectra have failed
due to the faintness of the H$\alpha$ emission line in
\object{NGC~3623} and to the low stellar velocity dispersion (narrow
H$\alpha$ absorption line) in \object{NGC~3627}. Ho et al.(\cite{hofil3})
subtracted a pure-absorption-line template from the observed spectra and 
obtained
$EW(\mbox{H}\alpha _{\mbox{emis}})=0.83$~\AA\ for \object{NGC~3623}
and 4.25~\AA\ for \object{NGC~3627}.
For radiative excitation of gas by OB-stars
(`\ion{H}{II}-region'-type excitation) the well-established and lowest
possible emission intensity ratio
H$\alpha /\mbox{H}\beta$ is 2.5--2.7, and it is much larger
for shock excitation of the gas. Ho et al. (\cite{hofil3}) have classified
the nuclear emission spectrum of \object{NGC~3623} as LINER and
that of \object{NGC~3627} as transition type between Seyfert 2
and \ion{H}{II}. Stasinska \&\ Sodre (\cite{sts2001}) have analysed
a large sample of integrated spectra of galaxies of various morphological
types and have found a good correlation 
$EW(\mbox{H}\beta _{\mbox{emis}})=0.25 EW(\mbox{H}\alpha _{\mbox{emis}})$;
this is just the relation we have to use to calculate 
$EW(\mbox{H}\beta _{\mbox{emis}})$,
which is in fact the correction of the Lick H$\beta$ index for the emission.
 We stress that any substantial contribution of an active nucleus
to the gas excitation near the centers of the galaxies or any dust extinction 
should increase the Balmer decrements and so diminish the correction for
the H$\beta$ emission estimated by our method; therefore the mean stellar
ages estimated below using the corrected and the uncorrected H$\beta$
values are indeed the lower and upper limits of the true luminosity-weighted
stellar ages.

\begin{figure}
 \resizebox{\hsize}{!}{\includegraphics{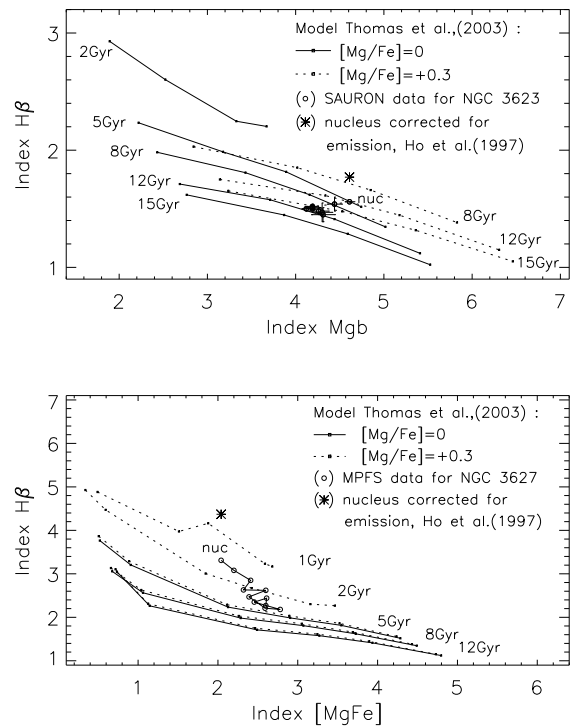}}
 \caption{Age-diagnostics diagrams:
 H$\beta$ vs Mgb for the azimuthally averaged  Lick indices in the center 
 of NGC~3623 and H$\beta$ vs [MgFe],
  [MgFe]$\equiv (\mbox{Mgb} <\mbox{Fe}>)^{1/2}$,
 for the azimuthally averaged  Lick indices in the center of NGC~3627.
 The galaxy data -- large signs connected by solid lines -- are taken 
 along the radius in steps of $\sim$1\arcsec; the nuclei are marked 
 `nuc'. The H$\beta$ measurements for the nuclei corrected for the
 emission contamination as described in the text are also plotted as
 separate asterisks.
 The models of Thomas et al. (2003) for [Mg/Fe]=0.0 and +0.3
 are plotted as a reference frame; the small signs connected by solid and
 dashed lines represent stellar population models of equal ages;
 the metallicities for the models are
 +0.67, +0.35, 0.0, -0.33, -1.35, and -2.25, if one takes the signs from
 right to left.}
 \label{ages}
\end{figure}

Figure~7 presents the age-diagnostic diagrams for our galaxies. For
\object{NGC~3623} (Fig.7 top) 
we compare H$\beta$ with Mgb because for this consideration we use
the SAURON data, and the SAURON iron index is strongly underestimated (see
above). Two models from Thomas et al. (\cite{thomod}) are plotted, for
[Mg/Fe]$=0.0$ and $+0.3$, because the abundance ratio in \object{NGC~3623}
is somewhere between these values, $+0.1 - +0.2$. The azimuthally averaged
`bulge' of \object{NGC~3623} is old, $T_{bul}=10-15$~Gyr, and has nearly
solar metallicity,
whereas the nucleus when corrected for the emission is found to have
an intermediate age, $\sim 5$~Gyr, and a very high metallicity, $+0.6$ dex.
These results are in general agreement with those of Proctor (\cite{prdiss}).
For \object{NGC~3627} (Fig.~7 bottom) we are able to compare H$\beta$ 
with the composite metal index
[MgFe]$\equiv \sqrt{\mbox{Mgb} \langle \mbox{Fe} \rangle}$, which 
eliminates completely the effect of the Mg/Fe ratio on age estimation. The
nucleus of \object{NGC~3627}  looks very young: after applying
the correction for the emission to the H$\beta$ index
we obtain $T\approx 1$~Gyr and nearly solar metallicity.
The bulge is found to have the same metallicity and an intermediate age
of the stellar population, $3-5$~Gyr. Note that in this age range
the possible systematic uncertainty of the H$\beta$ index by some 0.2~\AA\
which may be suspected for the MPFS data
has no substantial effect on the age estimates. The main conclusion of
this subsection is that the  mean stellar population properties
in the nuclei of \object{NGC~3623} and \object{NGC~3627} are quite different.

\section{Stellar and gas kinematics in the centers of NGC~3623 and NGC~3627
and their circumnuclear structures}

Since the integral-field spectroscopy provides us with two-dimensional
line-of-sight velocity fields, we are able now to analyse both
geometry of rotation and central structure of the galaxies.
If we have an axisymmetric mass distribution
and rotation on circular orbits, the direction of the maximum central
line-of-sight velocity gradient (we shall call it `kinematical major
axis') should coincide with the line of nodes, as should the
photometric major axis; whereas in the case of a triaxial potential
the isovelocities are aligned with the principal axis of the ellipsoid
 (see e.g. the simulations of bars by Vauterin \&\ Dejonghe
\cite{vd97} and references therein),
and generally the kinematical and photometric major axes diverge,
showing turns with respect to the line of nodes in opposite senses
if the main axis of the triaxial potential is not strictly aligned
with the line of nodes
(Monnet et al. \cite{mbe92}; Moiseev \&\ Mustsevoy \cite{mm2000}).
In a simple case of cylindric (disk-like) rotation we have a convenient
analytical expression for the azimuthal dependence of the central
line-of-sight velocity gradients within the area of solid-body rotation:\\
$$
dv_r/dr = \omega \sin i \cos (PA - PA_0),\eqno (1)
$$
where $\omega$ is the deprojected central angular rotation velocity,
$i$ is the inclination of the rotation plane, and $PA_0$ is the
orientation of the line of nodes, coinciding in the case of an
axisymmetric ellipsoid (or a thin disk) with the photometric
major axis. So by fitting the azimuthal variations
of the central line-of-sight velocity gradients by a cosine
law we can determine the orientation of the kinematical major
axis by its phase and the central angular rotation velocity
by its amplitude. It is our main tool of kinematical analysis.

\subsection{NGC 3623}

 de Zeeuw et al. (\cite{sau2}) have already claimed to detect the
kinematical signatures of the presence of a compact circumnuclear stellar
disk in the center of \object{NGC~3623} from their preliminary analysis of 
the SAURON data. Let us consider the issue in more detail.

\begin{figure*}
\centering
  \includegraphics[width=17cm]{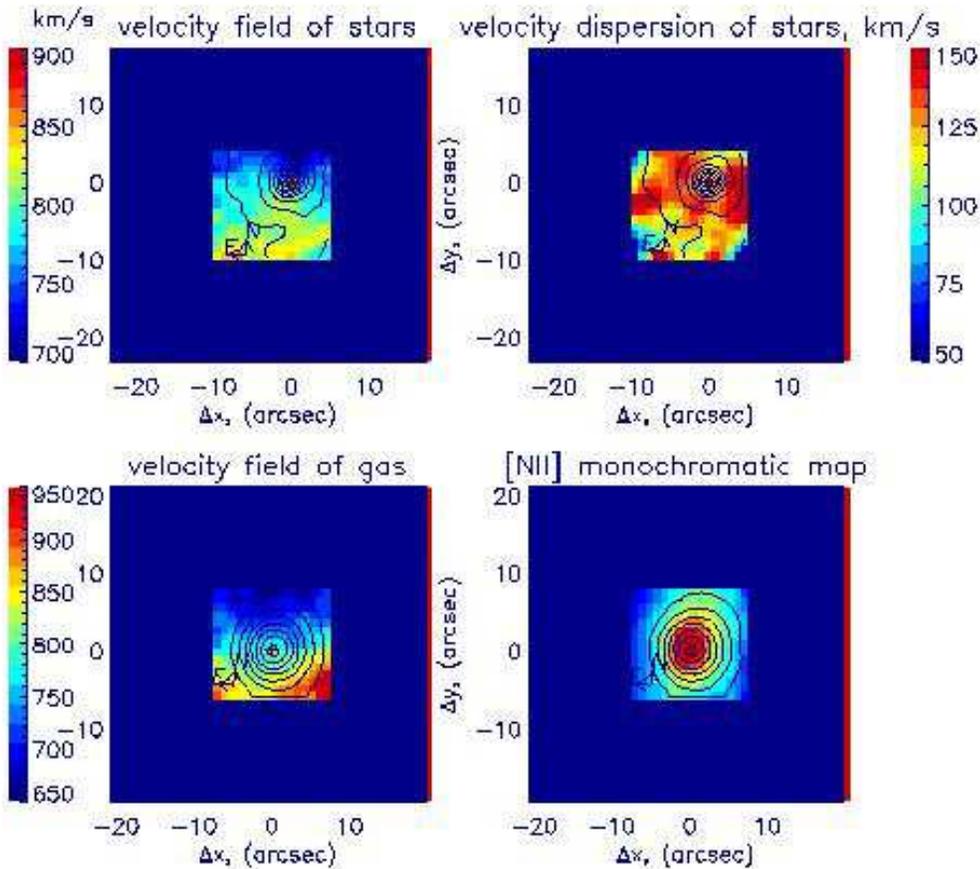}
  \caption{The line-of-sight velocity fields of the stellar component
  (top left, isolines) and of the ionized gas (bottom left, isolines),
  and the stellar velocity dispersion map (top right, gray-scaled) and the
  [NII] emission line intensity distribution (bottom right, gray-scaled)
  in the central part of NGC~3623 according to the MPFS data. 
  The continuum intensity is
  shown as gray-scaled in the left plots and as isophotes in the right plots.}
 \label{kin13623}
\end{figure*}

\begin{figure*}
\centering
  \includegraphics[width=17cm]{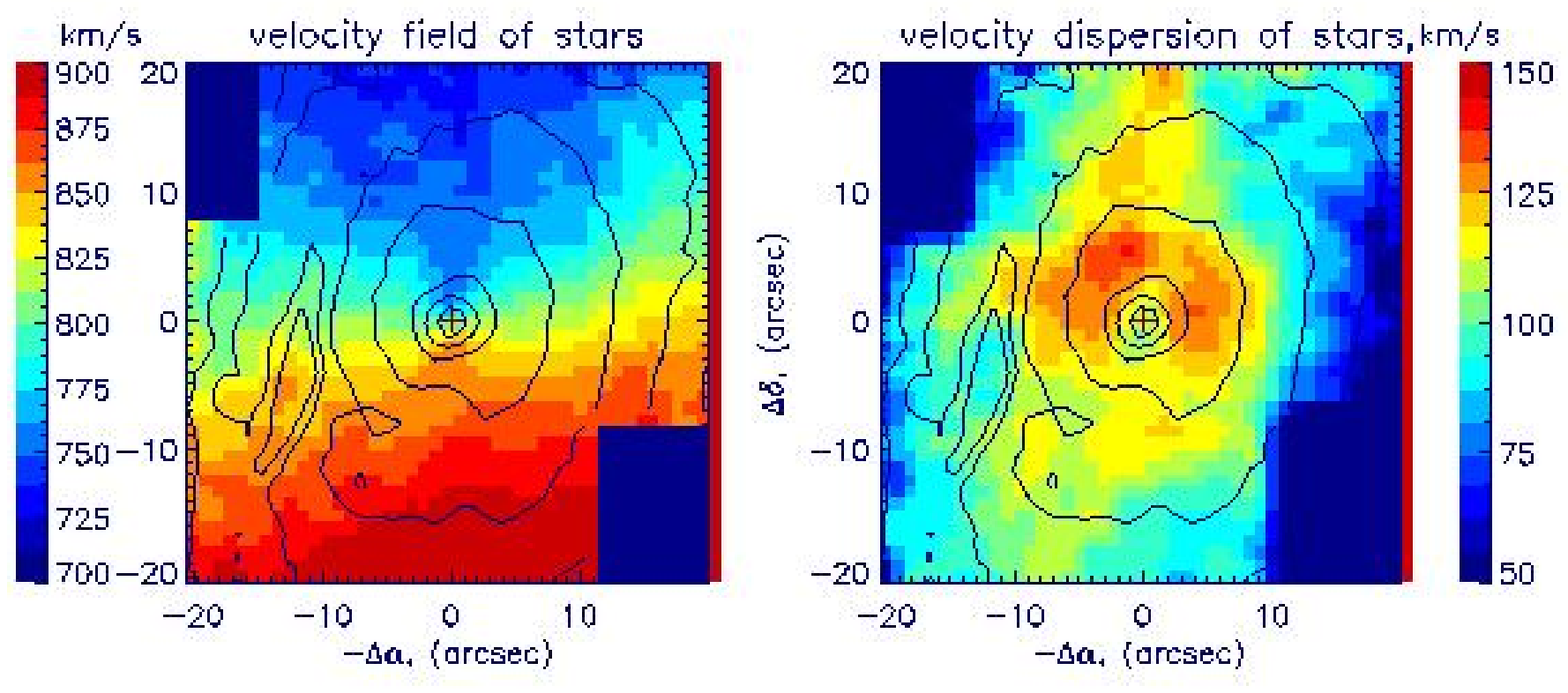}
  \caption{The line-of-sight velocity field of the stellar component
  (left, isolines) and the stellar velocity dispersion map (right, gray-scaled) 
  in the central part of NGC~3623 according to the SAURON data. 
  The continuum intensity is
  shown as gray-scaled in the left plot and as isophotes in the right plot.}
 \label{kin23623}
\end{figure*}

Figure~8 presents the kinematical information on \object{NGC~3623}
that we have derived
from the green and red MPFS spectra: line-of-sight velocity fields for the
stars and ionized gas, the stellar velocity dispersion field, and the 
[\ion{N}{II}]$\lambda$6583 emission-line intensity distribution
for comparison with
the orientation of the kinematical major axis of the ionized gas. Both
stellar and gas velocity fields within $R\le 5\arcsec$ imply regular rotation
in the symmetry plane of the galaxy whose line of nodes is at $PA_0=174\degr$
(see Table~1). Prominent crowding of the stellar isovelocities near the
major axis signifies the presence of a circumnuclear stellar disk; 
and the intensity distribution of the emission is
evidence for a gas confinement to the same circumnuclear disk. However,
a little
farther from the center the isovelocities, both of the stellar and gaseous
components, tend to turn consistently. This turn is much more evident in
Fig.~9, left, where the SAURON data combined from two different telescope
pointings are presented. Even the stellar isovelocities $\pm 40$ km/s from
the systemic velocity, which belong to the circumnuclear disk near the center,
certainly turn perpendicular to the north-north-east direction in the outer
parts of the field. Beyond $R\approx 8\arcsec$ we observe
practically cylindric rotation around a skewed rotation axis. The appearance
of the stellar velocity dispersion map is even more striking (Fig.~9,
right). There is a half-sphere of enhanced $\sigma _*$, with a radius
of $8\arcsec - 10\arcsec$, and with a hole inside it. 
The central minimum of the stellar velocity dispersion is also clearly seen 
in the MPFS data (Fig.~8) though the MPFS stellar velocity dispersion map
suggests a more complete ring of enhanced $\sigma _*$ closed
to the south of the nucleus. We should stress here that the zone of 
enhanced $\sigma _*$ coincides perfectly with the zone of low 
Mgb index (Fig.~2); both features are quite typical for a bulge. But it
 must mean again that the bulge of \object{NGC~3623} is confined within
$R=10\arcsec$ whereas from the analysis of photometric profiles it seems
to dominate within $R<20\arcsec$ (Burkhead \&\ Hutter \cite{burkphot}).
To check the bulge extension in \object{NGC~3623} we have undertaken
our own analysis of its surface brightness profile using the $K$-band
data of the 2MASS survey now available from NED; the results are
presented in Fig.~10. Indeed, inside
$R=30\arcsec$ the surface brightness profile can be perfectly fitted
by a de Vaucouleurs' law; moreover, the brightness profile of the
outer large-scale disk is of Freeman's II-type, that is, shows
a depression in the center. The inner truncation of the large-scale
stellar disk of \object{NGC~3623} at $R=98\arcsec$ has also been 
reported by Baggett et al. (\cite{bag}); they have found the effective radius
of the bulge in the $V$-band to be equal to $69\arcsec$. We cannot yet
understand this discrepance between the photometric size of the
bulge and its zone of influence on the dynamics and stellar
population properties integrated in the line of sight.

\begin{figure}
\resizebox{\hsize}{!}{\includegraphics{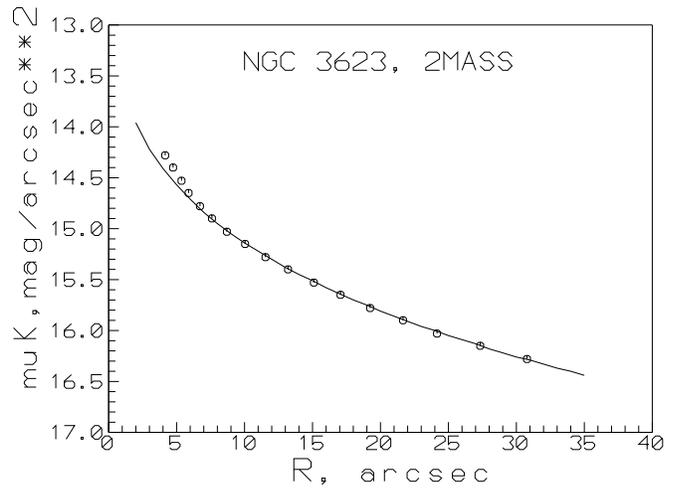}}
\caption{Fitting of the central azimuthally-averaged K-band brightness
profile of NGC~3623 by a de Vaucouleurs law. An excess of the brightness
within $R<6\arcsec$ may be attributed to the circumnuclear stellar
disk.}
\label{twomassprof}
\end{figure}

\begin{figure}
 \resizebox{\hsize}{!}{\includegraphics{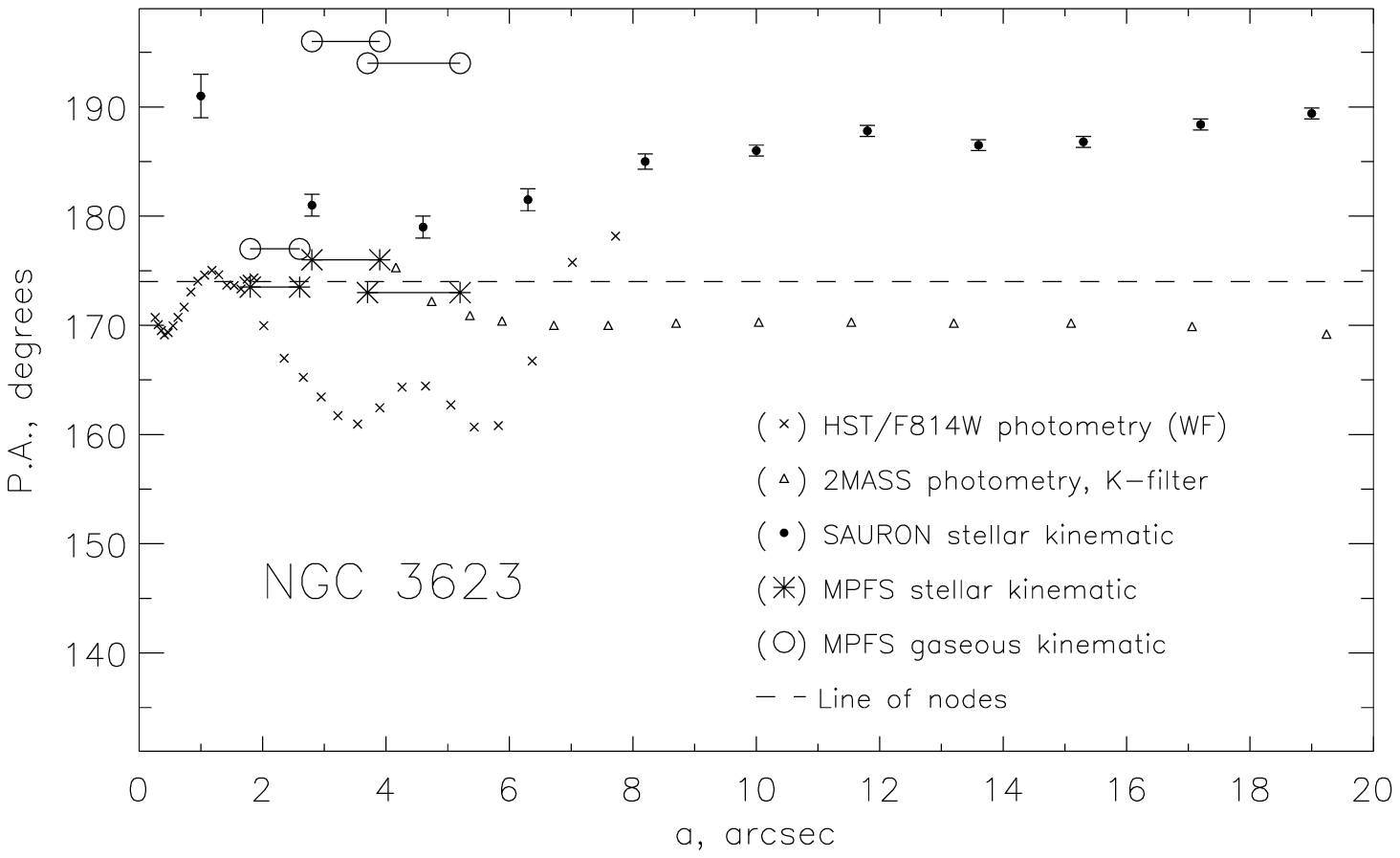}}
 \caption{Isophote major-axis position angle compared to the orientations of
    the kinematical major axes (see text) for the stars and ionized gas
    in the center of NGC~3623. The line of nodes determined
    from the outermost disk isophote orientation, $PA=174\degr$, is traced
    by a dashed line.}
 \label{iso3623}
\end{figure}

\begin{figure}
 \resizebox{\hsize}{!}{\includegraphics{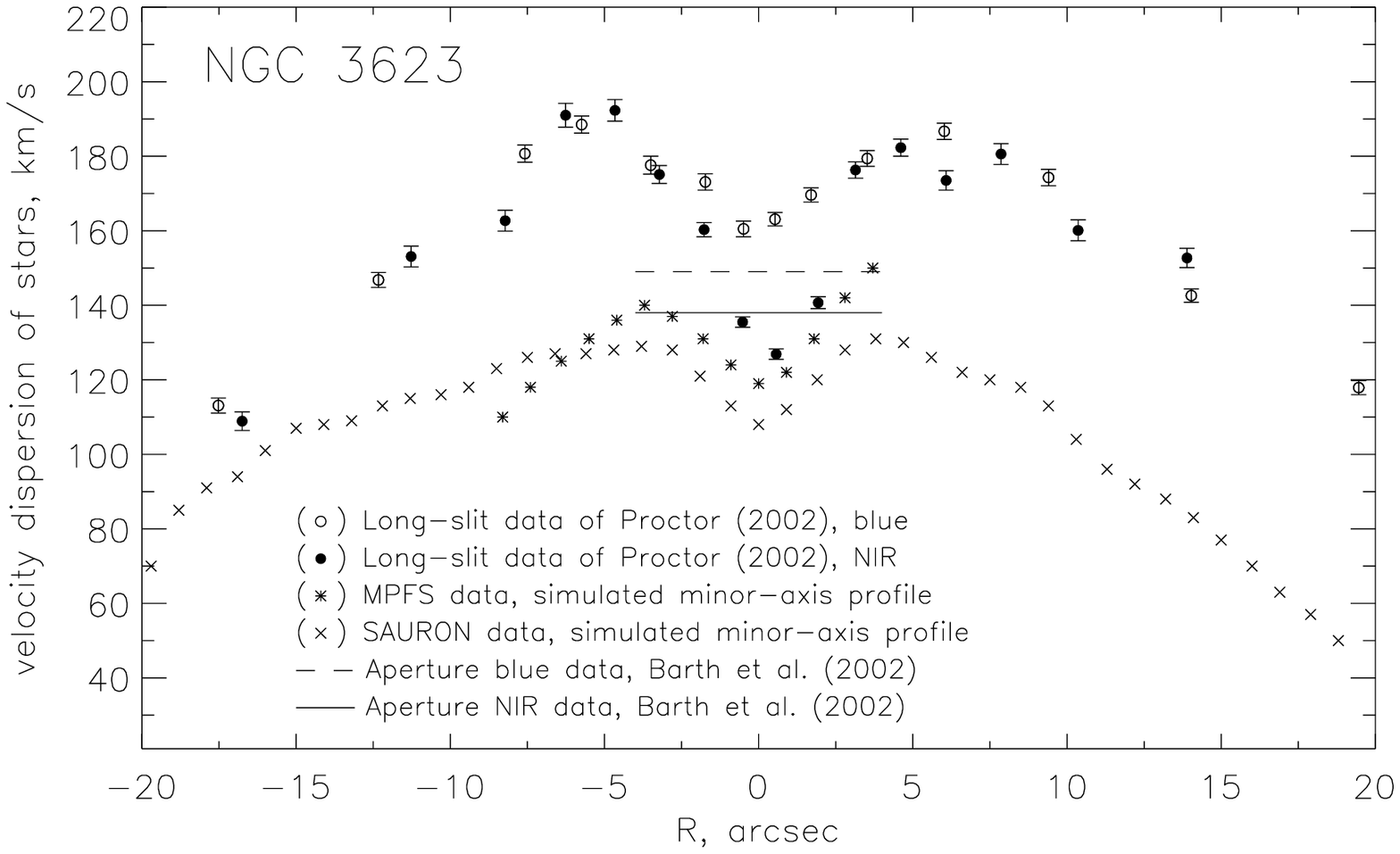}}
 \caption{Comparison of the stellar velocity dispersion profiles
 simulated along the minor axis using our 2D fields (MPFS and SAURON)
 for the center of NGC 3623 with the long-slit data of Proctor (2002) and
 aperture data of Barth et al. (2002).
  The slit width used in the simulations is 1.25\arcsec.
 }
\label{cuts3623}
\end{figure}

Figures~\ref{iso3623} and \ref{cuts3623} quantify our impressions
about stellar and gaseous kinematics. In Fig.~\ref{iso3623}
we compare the orientations of the kinematical major axes with those of the
isophotes. In the very center, $R\le 2\arcsec$, the HST/WFPC2/F814W
data reveal the isophote orientation coinciding with the outer line of
nodes. The drop of $PA$  at larger $R$ is perhaps an artifact of a dust effect
because it is not supported by the low-resolution K-data of the 2MASS survey.
The kinematical major axis orientations obtained from the MPFS data by
fitting formula (1) to the azimuthal variations of the line-of-sight
velocity gradients imply that the stars rotate axisymmetrically
in the plane of the circumnuclear disk (now definitely a disk, not
a bar) up to $R\approx 6\arcsec$ whereas the gas, even rather close to
the center, $R\ge 3\arcsec$, shows some signs of non-circular
motions. The SAURON stellar velocity field has been analysed by applying
the well-known method of tilted rings (Begeman \cite{begeman}) realized
by Dr. A.V.~Moiseev under IDL as the software DETKA. The SAURON data
reveal a turn of the stellar kinematical major axis at 
$R=8\arcsec - 20\arcsec$ by $+10\degr - +15\degr$ from the line of nodes;
as the large bar of \object{NGC~3623} is elongated in $PA\approx 155\degr$,
that is, its $\Delta PA\approx -19\degr$, the whole picture is consistent
with the stellar elliptical x1 orbits within the bar potential. It is an
additional proof of what we have already established from the index and
$\sigma _*$ maps: that at $R>8\arcsec$ we do not see noticeable  
influence of the bulge stellar component.

The minimum of the stellar velocity dispersion in the nucleus of
\object{NGC~3623} has already been reported by Proctor et al.
(\cite{prsanr00}) and by Proctor (\cite{prdiss})
from their long-slit observations along the minor axis of the galaxy.
In Fig.~\ref{cuts3623} we compare the data of Proctor (\cite{prdiss})
with the minor-axis
cross-sections simulated over the MPFS and SAURON stellar velocity dispersion
maps. Although there is a noticeable systematic shift between our measurements
and those of Proctor (\cite{prdiss}), the qualitative behaviour of 
the radial profiles
of the stellar velocity dispersion is the same: the highest $\sigma _*$ is
observed at $R=4\arcsec - 6\arcsec$, and in the nucleus there is a drop
of $\sigma _*$ by some 20\%. It is interesting that according to
Proctor (\cite{prdiss}), who has measured the stellar velocity dispersion
using two different spectral ranges, in the NIR \ion{Ca}{II} triplet
the drop of $\sigma_*$ is deeper and reaches about 30\%; it may be
an evolutionary effect of probing stars of different ages in different
spectral ranges. Of course, we wonder which scale of $\sigma _*$ is
correct, ours or that of Proctor (\cite{prdiss}).
In Fig.~\ref{cuts3623} we have also plotted the quite
recent aperture measurements of Barth et al. (\cite{bhsdisp}), also
obtained in two spectral ranges. The latter data imply that 
`the truth is somewhat between' our and Proctor's measurements.

\subsection{NGC 3627}

\begin{figure*}
\centering
  \includegraphics[width=17cm]{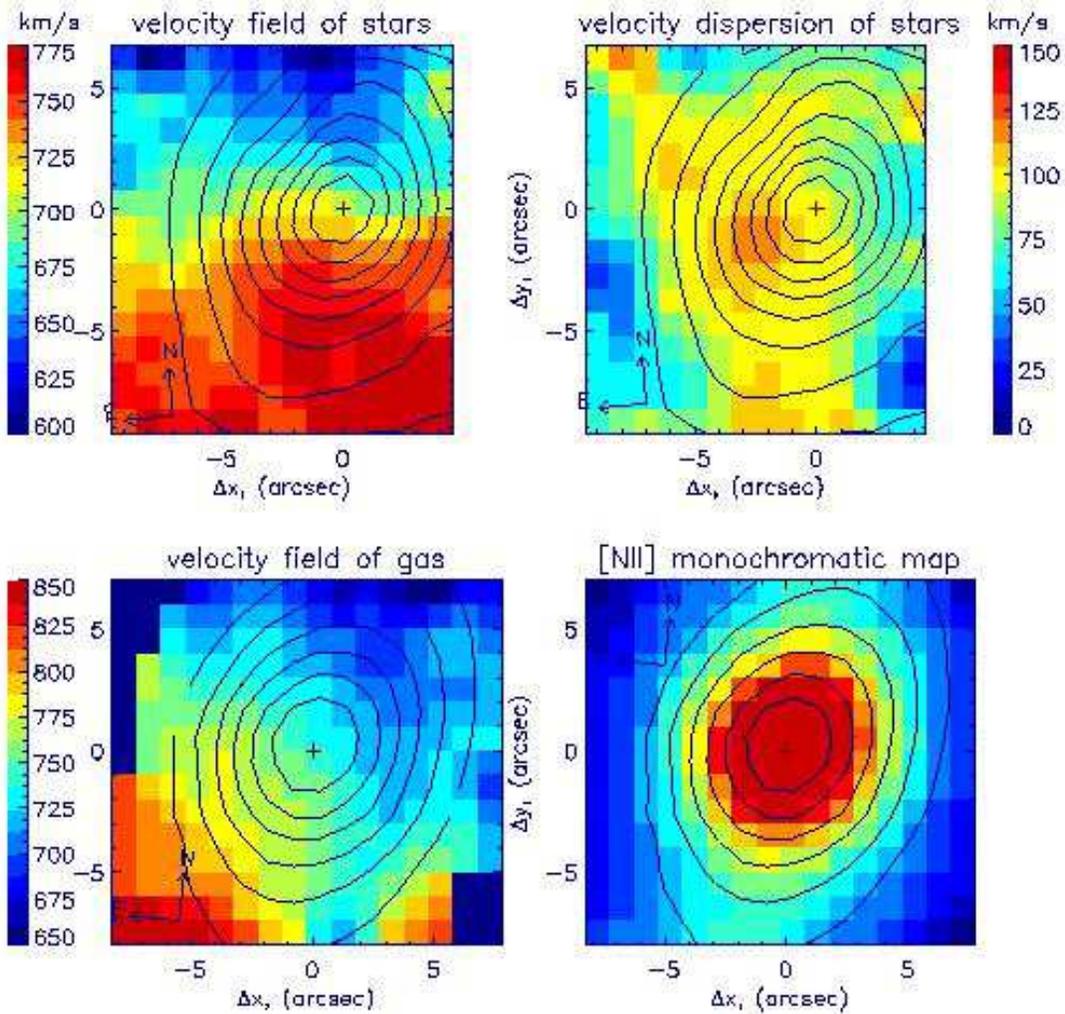}
  \caption{The line-of-sight velocity fields of the stellar component
  (top left, isolines) and of the ionized gas (bottom left, isolines),
  and the stellar velocity dispersion map (top right, gray-scaled) and the
  [NII] emission line intensity distribution (bottom right, gray-scaled)
  in the central part of NGC~3627. The continuum intensity is
  shown as gray-scaled in the left plots and as isophotes in the right plots.}
 \label{kin3627}
\end{figure*}

Figure~\ref{kin3627} presents kinematical maps for the center of
\object{NGC~3627}. Although
the spatial resolution of both the green and the red observations is rather
low (see Table~2) and we cannot trace any variations of the rotation 
over the field of view, its total character is clear. The stellar
component rotates regularly and axisymmetrically and probably forms a
circumnuclear disk or an oblate bulge. As for the ionized gas, its
emission brightness distribution is a twin to that in the center of
\object{NGC~3623}, but the velocity field is quite different: all
the isovelocities,
from the innermost to the outermost ones, are turned  as a whole with
respect to the kinematical major axis of the stars. Finally, the
stellar velocity dispersion distribution shows a peak instead
of the hole of \object{NGC~3623}; however, this peak does not coincide exactly
with the continuum peak, and it is shifted in another direction from
the position of the H$\beta$ index peak.

\begin{figure}
 \resizebox{\hsize}{!}{\includegraphics{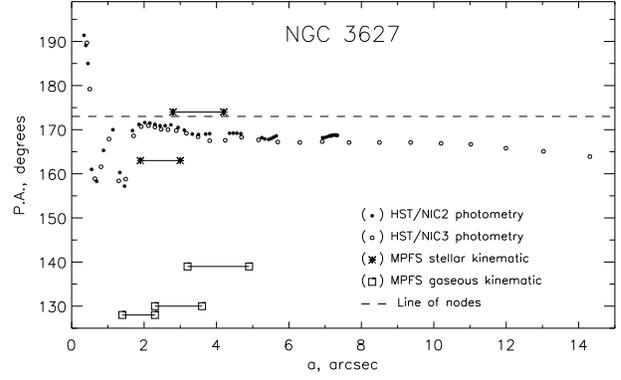}}
 \caption{Isophote major-axis position angle compared to the orientations of
    the kinematical major axes (see text) for the stars and ionized gas
    in the center of NGC~3627. The line of nodes determined
    from the outermost disk orientation, $PA=173\degr$, is traced
    by a dashed line.}
 \label{iso3627}
\end{figure}

In Fig.~\ref{iso3627} we compare the orientations of the kinematical major
axes with the orientation of the NIR (HST/NICMOS) central isophotes. Taking
into account the difference of spatial resolution, we may conclude that the
stellar rotation follows the orientation of the photometrical major axis,
and at $R=2\arcsec - 5\arcsec$ both are close to the line of nodes of the
global disk of \object{NGC~3627}.
This double coincidence implies an axisymmetry
of the central part of the galaxy. The kinematical major axis of the
ionized gas deviates by $-35\degr$ to $-40\degr$ from the line of nodes and
from the elongation of the emission brightness distribution (see 
Fig.~\ref{kin3627}). Since the global bar of \object{NGC~3627} is aligned
at $PA\approx 160\degr$, or deviates from the line of nodes in the same
negative sense of $\Delta PA$, the orientation of the gas kinematical
major axis cannot be explained by elliptical x1 orbits, as 
has been done in the case of \object{NGC~3623}. Since the morphology
of the gas and/or dust distribution in the center of \object{NGC~3627}
does not imply any regular off-plane structures like disks or rings, the
most natural way to explain the gas velocity field of \object{NGC~3627}
may be by strong radial motions. If we suppose that the
eastern side of the galaxy is the nearest to us by assuming that its
spiral arms are trailing, we should conclude that perhaps
we deal with gas inflow into the nucleus.

\section{Conclusions and Discussion}

Unlike our previous targets, \object{NGC~3384}/\object{NGC~3368}
in the Leo I Group and \object{NGC~5574}/\object{NGC~5576} in LGG~379,
\object{NGC~3623} and \object{NGC~3627} do not appear to
experience synchronous evolution. The mean ages of their circumnuclear
stellar populations are quite different. Since to derive the stellar
population properties we have used the model synthesis calculations for
Simple Stellar Populations (SSPs), that is, for stellar systems with
homogeneous ages and metallicities of stars, the results are somewhat
conventional: the real galaxies consist of stars of various ages.
However due to fast luminosity evolution of stellar populations our
luminosity-weighted mean age estimate must be close to an age of the
latest star formation event. The magnesium overabundance of the nucleus
in \object{NGC~3627} found here is evidence for a very brief last star
formation event $\sim 1$ Gyr ago whereas the central part
of \object{NGC~3623} looks quiescent. Moreover, though the global bars
of both galaxies have similar morphologies, it is only in the center of
\object{NGC~3627} that we observe noticeable irregular, possibly radial
motions of the gas, and the stars
and the ionized gas in the center of \object{NGC~3623} show more
or less stable rotation. However, \object{NGC~3623} has a chemically
distinct core -- a relic of a circumnuclear star formation
burst -- which is shaped as a compact, dynamically cold stellar disk with
a radius of $\sim 250-350$ pc that has been formed not later that 5 Gyr ago.

What may be the reason for these different properties of the central regions 
in  \object{NGC~3623} and \object{NGC~3627}? Perhaps just what was guessed
by Zwicky (\cite{zwicky}) and Vorontsov-Vel'yaminov (\cite{vv_p1}):
the latter is an interacting galaxy and the former is not.
Rots (\cite{rots78}) tried to explain a spectacular long plume of
\object{NGC~3628} seen both in the optical image and in the 21 cm neutral 
hydrogen line 
and proposed a scenario of tidal interaction with \object{NGC~3627}.
In the frame of his model \object{NGC~3627}, moving along a parabolic orbit,
passed through perigalacticon at 25 kpc from \object{NGC~3628} some
$8 \times 10^8$ yr ago. Though the numerical simulations were very crude
-- \object{NGC~3628} consisted of only 441 test particles and
\object{NGC~3627} was taken as a single gravitating point, --
Rots (\cite{rots78}) obtained a picture quite consistent with the morphology 
of \object{NGC~3628}. He argued that \object{NGC~3627} could not experience
a noticeable tidal transformation
during this encounter because its orbit was almost retrograde with respect
to its intrinsic rotation. However, Zhang et al. (\cite{zhang}) 
combined high-resolution observations of \object{NGC~3627}
in CO and in \ion{H}{I} and found certain signatures of violent dynamical
evolution of the galaxy after the proposed encounter with \object{NGC~3628}.
First of all, the outermost southern \ion{H}{I} clouds show
visible counterrotation with respect to the more inner gas. But what
is even more striking is an enormous gas concentration in the center
of the galaxy. Zhang et al. (\cite{zhang}) detect an inner molecular-gas
bar with total mass not less than $4 \times 10^8 \, M_{\odot}$;
this means that according to their estimates the ISM contributes not less than
30\%\ to the total dynamical mass within $R=750$ pc. For comparison, within
$R=7$ kpc the ISM mass constitutes less than 3\%\ of the dynamical mass.
Zhang et al. (\cite{zhang}) conclude that
`continuous and efficient radial mass accretion across the entire galactic disk'
is unavoidable in \object{NGC~3627}. The signatures of this accretion may be 
directly observed by us in the central gas velocity field
as the turn of the kinematical major axis of the gas. Note also that the time
passed since the close encounter with \object{NGC~3628} in the simulations
by Rots (\cite{rots78}) coincides with the age of the
last star formation burst in the nucleus of \object{NGC~3627}. Was this
burst triggered by the tidal interaction? And the last question inspired
by the successful model of Rots (\cite{rots78}): if the whole picture
requires a parabolic orbit, was \object{NGC~3627} a member of the
\object{Leo Triplet} before its encounter with \object{NGC~3628}?
If not, the visible resemblance between \object{NGC~3623}
and \object{NGC~3627} is a simply chance, and we should not expect
synchronous evolution of
\object{NGC~3623} and \object{NGC~3627} over timescales of several Gyrs.

Another interesting problem relates to the drop in the stellar velocity 
dispersion 
in the nucleus of \object{NGC~3623}. There are several cases known of
stellar velocity dispersion drops in the centers of early-type disk galaxies,
and recently Wozniak et al. (\cite{dispdrop}) have tried to model this
phenomenon in the frame of
self-consistent N-body simulations including stars, gas, and star formation.
Indeed, after $\sim 0.4$ Gyr of numerical evolution their galaxy has developed
a central velocity dispersion drop due to nuclear gas inflow, subsequent
star formation and the appearance of young luminous stars born from dynamically
cold gas. However, the dynamical evolution seems to be rather fast, and after
another 0.5~Gyr the central $\sigma _z$ of the `new stellar population'
increases by a factor of 2 due to the gravitational interaction with the gas
and with the old and hotter population. To put this into perspective, 
Wozniak et al. (\cite{dispdrop}) note that 
`$\sigma _z$ of the new population is expected
to become similar to the $\sigma _z$ of the old population in at least 2~Gyr'.
Let us compare these model timescales to our estimates of ages of the
stellar
populations in the centers of \object{NGC~3623} and \object{NGC~3627}.
\object{NGC~3623} has an age of the nuclear
stellar population not less than  5~Gyr and a prominent stellar velocity
dispersion drop in the nucleus. Much younger stars dominate in the
nucleus of \object{NGC~3627},
their $T\sim 1$ Gyr, and it has only marginal $\sigma _*$
depression near the center, according to the long-slit data of Heraudeau
\&\ Simien (\cite{hs98}); in our panoramic data 
we see not even a marginal $\sigma _*$ drop in the center 
of this galaxy but rather a peak in the stellar velocity dispersion shifted 
slightly with respect to the photometric center. The relation
between the age of the nuclear stellar population and the prominence of the
drop in the stellar velocity dispersion obtained by us is in fact
opposite to the model predictions by Wozniak et al. (\cite{dispdrop}).
Moreover, recently we found a strong example of a stellar velocity
dispersion drop in a galaxy center where the mean stellar age is older
than 12 Gyr -- it is NGC~2768 (Sil'chenko \&\ Afanasiev \cite{we2004};
see also Emsellem et al. \cite{sau3}).
So, though the statistics is rather small yet, no anti-correlation between
the prominence of the drop in the stellar velocity dispersion and 
the age of the stellar population in the center of a galaxy is seen.
We have to admit that the problem of the origin of the drop in velocity 
dispersion in the center and of its long life remains unsolved.

\begin{acknowledgements}
We thank Dr. A. V. Moiseev of SAO RAS for analysing the stellar
velocity field of \object{NGC~3623} with his software DETKA.
During the data analysis we have
used the Lyon-Meudon Extragalactic Database (LEDA) supplied by the
LEDA team at the CRAL-Observatoire de Lyon (France) and the
NASA/IPAC Extragalactic Database (NED) which is operated by the
Jet Propulsion Laboratory, California Institute of Technology,
under contract with the National Aeronautics and Space Administration.
This research is partly based on data taken from the ING Archive
of the UK Astronomy Data Centre and
on observations made with the NASA/ESA Hubble Space Telescope, obtained
from the data archive at the Space Telescope Science Institute, which is
operated by the Association of Universities for Research in Astronomy,
Inc., under NASA contract NAS 5-26555.
The study of nearby galaxy groups
is supported by the grant of the Russian Foundation for Basic Researches
04-02-16087 and the study of the evolution of galactic centers --
by the Federal Scientific-Technical Program -- contract
of the Science Ministry of Russia no.40.022.1.1.1101.

\end{acknowledgements}

\end{document}